\documentclass{article}
\usepackage{arxiv}
\usepackage[utf8]{inputenc} 
\usepackage[T1]{fontenc}    
\usepackage[colorlinks,linkcolor=blue]{hyperref}       
\usepackage{url}            
\usepackage{booktabs}       
\usepackage{nicefrac}       
\usepackage{microtype}      
\usepackage{lipsum}		
\usepackage{graphicx}
\usepackage{doi}
\usepackage{bm}
\usepackage{amsmath,amssymb,amsfonts}
\usepackage{algorithmic}
\usepackage{graphicx}
\usepackage{textcomp}
\usepackage{xcolor}
\usepackage{multirow}

\title{Deep Geometric Distillation Network for Compressive Sensing MRI\thanks{Accepted by IEEE-EMBS International Conference on Biomedical and Health Informatics (BHI), 2021.}}

\author{ \hspace{1mm}Xiaohong Fan, 
\hspace{1mm}Yin Yang,
\hspace{1mm}Jianping Zhang\thanks{Corresponding author: \texttt{jpzhang@xtu.edu.cn} (J. Zhang).}
 \\
	School of mathematics and computational science\\ Xiangtan University\\
	Xiangtan, P. R. China \\
	\texttt{fanxiaohong@smail.xtu.edu.cn, \{yangyinxtu,jpzhang\}@xtu.edu.cn}\\
}

\renewcommand{\shorttitle}{Deep Geometric Distillation Network for CS-MRI}

\begin{document}
\maketitle

\begin{abstract}
Compressed sensing (CS) is an efficient method to
reconstruct MR image from small sampled data in $k$-space and accelerate the acquisition of MRI. In this work, we propose a novel deep geometric distillation network which combines the merits of model-based and deep learning-based CS-MRI methods, it can be theoretically guaranteed to improve geometric texture details of a linear reconstruction. Firstly, we unfold the model-based CS-MRI optimization problem into two sub-problems that consist of image linear approximation and image geometric compensation. Secondly, geometric compensation sub-problem for distilling lost texture details in approximation stage can be expanded by Taylor expansion to design a geometric distillation module fusing features of different geometric characteristic domains.
Additionally, we use a learnable version with adaptive initialization
of the step-length parameter, which allows model more flexibility that
can lead to convergent smoothly.
Numerical experiments verify its superiority over other state-of-the-art CS-MRI reconstruction approaches. The source code will be available at \url{https://github.com/fanxiaohong/Deep-Geometric-Distillation-Network-for-CS-MRI}.
\end{abstract}


\section{Introduction}
Magnetic Resonance Imaging (MRI) is a widely used medical imaging modality for clinical diagnosis. Nyquist sampling theory has guided the signal sampling process for many years. Strict sampling requirements put a great burden on processing equipment. CS-MRI \cite{Donoho2006} applies the sparse characteristics of signals in $k$-space to image reconstruction, and the sampling rate is much lower than Nyquist sampling rate, but the image quality is not significantly reduced.

Classical CS-MRI methods can be explored in a specific transformation domain or in a generic dictionary-based subspace to learn more flexible sparse representation directly from data \cite{Lustig2007}. Although Total Variation (TV) regularization in the gradient domain introduces staircase artifacts in reconstructed image, it has been widely used in MRI reconstruction due to simple and fast \cite{Block2007}. Discrete wavelet transform \cite{Lai2016} and discrete cosine transform \cite{Lingala2013} have also been utilized to reconstruct CS-MRI. Many iterative algorithms such as Iterative Shrinkage Threshold Algorithm (ISTA) \cite{Beck2009} and Alternating Direction Multiplier (ADMM) \cite{Boyd2010} have been designed to solve CS-MRI models. These CS-MRI methods are all based on interpretable and predefined sparsity image prior rather than direct learning, but cannot obtain desirable reconstruction results. Although most of these methods have the advantages of theoretical analysis and strong convergence, they usually require expensive computational complexity to converge, and also face the difficulties of selecting the regularizer and model parameter.

Recently, deep learning has received great success in computer vision community. CS-MRI reconstruction has been solved by a generalized inverse problem such as image super-resolution (SR) \cite{Dong2016} and denoising \cite{Agostinelli2013}. A multi-layer Convolutional Neural Network (CNN) has been proposed to recover fully sampled MRI from under-sampled MRI \cite{Wang2016b}.
These existing learning-based CS-MRI methods are driven by a large amount of training samples as a black box without model prior.

Deep unrolling method combines the advantages of model-based and learning-based CS-MRI methods, which has sufficient theoretical support and good performances \cite{yangyan2016}. ISTA is mapped into deep CNN network to learn proximal mapping \cite{Zhang2018}. FISTA-Net is designed by mapping FISTA \cite{Lustig2007} algorithm into a deep network that consists of three update blocks, i.e. gradient descent, proximal mapping and acceleration \cite{Xiang2020}.

However, ISTA-Net+ and FISTA-Net doesn't make full use of other regularization priories (just $\ell_1$ prior information), and the relationship from optimization theory to network design is also not natural enough. Based on these drawbacks, we start from CS-MRI problem and transform it into a problem that distills features of different geometric characteristic domains and naturally corresponds to the design of our network. The main contributions of this work can be summarized as follows
\begin{itemize}
\item[(1)] We propose a novel deep geometric distillation network, which can not only be theoretically guaranteed to improve partial geometric texture details, but also give us a new perspective to design network structures.
\item[(2)] We compensate the lost texture information from different geometric characteristic domains by designing a geometric distillation module, which is inspired by Taylor expansion of the nonlinear geometric operator.
\item[(3)] All parameters in the proposed method
are learnable and these designed constraints can be used to ensure converging smoothly.
\end{itemize}

\section{Methodology}
\label{sec:headings}
\begin{figure*}[h]
\centerline{\includegraphics[width=0.98\columnwidth]{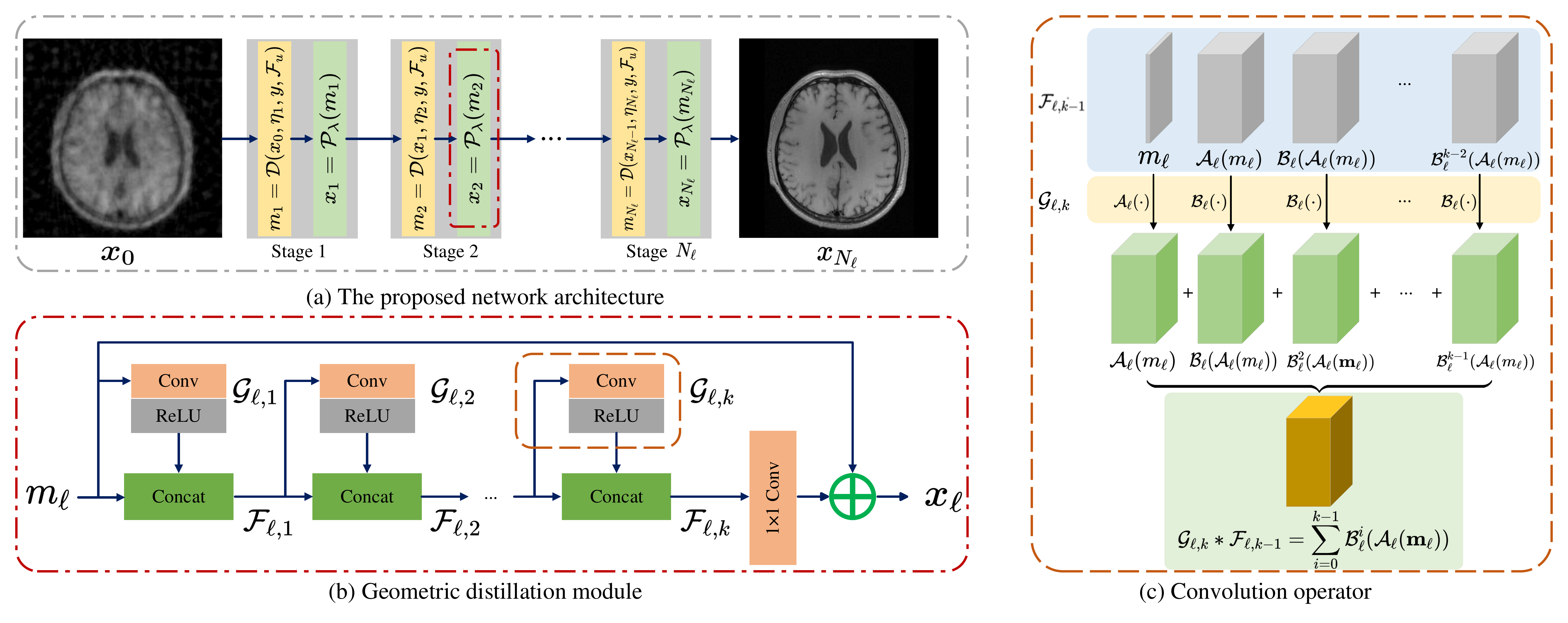}}
\caption{The overall architecture of deep geometric distillation network for Compressive Sensing MRI reconstruction.}
\label{fig1}
\end{figure*}

The goal of this work is to achieve MRI reconstruction
based on Compressive Sensing theory and deep CNN framework.
The overall architecture of our deep geometric distillation network is shown in Fig.\ref{fig1}, and more details are provided hereafter.

\subsection{Iterative CS-MRI Reconstruction}
General CS-MRI reconstruction can be formulated as the following optimization problem
\begin{equation}
\min_{\bm{x}}\left\{\mathcal{E}(\bm{x}):= \mathcal{S}(\bm{x})+\gamma \mathcal{R}(\bm{x})\right\},
\label{eq1}
\end{equation}
where $\bm{x} \in \mathbb{R}^{N}$ represents a reconstruction of the target MRI, $\mathcal{S}(\bm{x})=\frac{1}{2}\left\|\mathcal{F} \bm{x}-\bm{y}\right\|_{2}^{2}$ is a data fidelity term and $\mathcal{R}(\bm{x})$ is a regularizer with data prior, $\bm{y} \in \mathbb{R}^{M} (M\ll N)$ represents the under-sampled $k$-space data, $\mathcal{F}\in \mathbb{R}^{M\times N} $ is an under-sampled Fourier measurement matrix, $\gamma$ is a regularization parameter. Moreover, for iteration stage $\ell$, $\mathcal{S}_\ell(\bm{x})$ is the second-order Taylor expansion of $\mathcal{S}(\bm{x})$ at $\bm{x}_{\ell-1}$, which is denoted by
\[\mathcal{S}(\bm{x})\approx\mathcal{S}_\ell(\bm{x})=\frac{1}{2\eta_{\ell}}\|\bm{x}-(\bm{x}_{\ell-1}-\eta_{\ell}\mathcal{F}^{T}(\mathcal{F} \bm{x}_{\ell-1}-\bm{y}))\|_{2}^{2}+c,\]
hence we can obtain the solution of \eqref{eq1} via two-step iterations
\begin{align}
\bm{m}_{\ell}&=\mathcal{D}(\bm{x}_{\ell-1},\eta_{\ell},\bm{y},\mathcal{F}):=\underset{\bm{x}}{\arg \min }\;\mathcal{S}_\ell(\bm{x}),  \label{eq10}\\
\bm{x}_{\ell}&=\mathcal{P}_\lambda(\bm{m}_\ell):=\underset{\bm{x}}{\arg \min } \left\{\frac{1}{2}\|\bm{x}-\bm{m}_{\ell}\|_{2}^{2}+\lambda_\ell\mathcal{R}(\bm{x})\right\}
\label{eq11}
\end{align}
where $1/\eta_{\ell}$ is a Lipschitz constant, $c$ is a constant w.r.t $\bm{x}_{\ell-1}$ and $\lambda_\ell=\eta_{\ell}\gamma$.

\subsection{Deep Geometric Distillation Network}
One alternative to the above iterative optimization scheme is to perform a linear reconstruction module $\mathcal{D}$
followed by a learned feature compensation module $\mathcal{P}_\lambda$. The module $\mathcal{D}$ often result in heavy artifacts, while the designed CNN module $\mathcal{P}_\lambda$ is learned to compensate more texture details.

\textbf{Linear reconstruction module $\mathcal{D}$:}
Most methods to reconstruct an image from its measurements $\bm{y}$ rely mainly on
linear reconstruction update (\ref{eq10}) with a step-length parameter $\eta_{\ell}$ and a deterministic operator $\mathcal{F}$. In any practical setting, to increase network flexibility, the step size $\eta_{\ell}$ should be positive and decrease with the increasing of iterations smoothly. There are a variety of ways to use training data to adaptively learn
parameters $\eta_{\ell}$. Here we employ the \textbf{softplus} function $\bm{sp}(z)=\ln (1+\exp (z))$ \cite{Xiang2020} to make the proposed network converge properly. Mathematically, we rewritten \eqref{eq10} as
\begin{equation}
\bm{m}_{\ell}=\mathcal{D}(\bm{x}_{\ell-1},\eta_{\ell},\bm{y},\mathcal{F})=\bm{x}_{\ell-1}-\eta_{\ell}\mathcal{F}^{T}(\mathcal{F} \bm{x}_{\ell-1}-\bm{y})
\label{eq33}
\end{equation}
with a learnable step-length $\eta_{\ell}$, and its initial guess is given as follow
\begin{equation}
\eta_{\ell}=\bm{sp}\left(c_{1} \ell+c_{2}\right), \; c_{1}<0,\; \ell=1,2, \ldots, \mathcal{N}_{\ell}.
\label{eq32}
\end{equation}

\textbf{Geometric distillation module $\mathcal{P}_\lambda$}:
On the basis of data module $\mathcal{D}$, we further propose a prior feature
compensation module to compensate for lost texture details. By partially distilling the components from output $\bm{m}_{\ell}$ of data module $\mathcal{D}$, we can obtain feature maps originating from different sparse domains in different stages. Then, these features are aggregated into $\bm{m}_{\ell}$ to purify and gain more abundant and efficient geometric information.

We recall the optimal condition of \eqref{eq11} as
\[\bm{x}_{\ell}-\mathcal{W}(\bm{x}_{\ell})=\bm{m}_{\ell},\]
where $\mathcal{W}(\bm{x})=-\lambda_\ell\frac{\partial \mathcal{R}(\bm{x})}{\partial\bm{x}}$ represents the geometric characteristics of $\bm{x}$.
Since $\mathcal{W}(\bm{x})$ is non-linear, it is difficult to directly obtain the close-form solution $\bm{x}_{\ell}=(I-\mathcal{W})^{-1}(\bm{m}_{\ell})$ from an input $\bm{m}_{\ell}$. Naturally, if the operator $\mathcal{W}$ satisfies the spectral constraint $\|\mathcal{W}\|<1$, thus we can simplify operator $(I-\mathcal{W})^{-1}$ by using Taylor expansion as follow:
\begin{align}
\begin{split}
\bm{x}_{\ell}&=(I-\mathcal{W})^{-1}(\bm{m}_{\ell})=\left(\sum_{i=0}^{n}\mathcal{W}^{i}+R(\mathcal{W}^{n})\right)(\bm{m}_{\ell})\\
&\in\text{span}(\bm{m}_{\ell},\mathcal{W}(\bm{m}_{\ell}),...\,\mathcal{W}^{n}(\bm{m}_{\ell}), R(\mathcal{W}^{n})),
\end{split}
\label{eq15}
\end{align}
where $R(\mathcal{W}^{n})$ is the remainder of Taylor's expansion and is replaced by $\mathcal{W}^{n+1}(\bm{m}_{\ell})$.

The geometric texture information in different sparse (geometric) domains can be represented by the partial derivatives of MRI image $\bm{x}$. Fortunately, the convolution in neural network can be seen as the combination of several derivative operations.
Next we hope to design a network operator $\mathcal{G}_{\ell}(\bm{m}_{\ell})$ (or $\mathcal{G}_{\ell,k}$) replacing $\mathcal{W}$ for compensating lost geometric characteristics of $\bm{x}_{\ell}$. This approach is advantageous because the learning of
$\mathcal{G}_{\ell,k}$ has a large impact on the quality of the reconstruction.

Let $\text{Concat}(\cdot,\cdot)$ be a concatenate operator as follow
\[\text{Concat}(\bm{f}_1,\bm{f}_2)=(\bm{f}_1,\bm{f}_2)\in\mathbb{R}^{n\times m\times s}\]
where $\bm{f}_1\in\mathbb{R}^{n\times m\times s_1}$, $\bm{f}_2\in\mathbb{R}^{n\times m\times s_2}$ and $s=s_1+s_2$. Let $\mathcal{F}_{\ell,i}$ be the feature map of the layer $i$ in stage $\ell$ defined by
 \begin{equation*}
 \begin{split}
 \mathcal{F}_{\ell,0}&=\bm{m}_{\ell},\;\mathcal{F}_{\ell,1}=\big(\bm{m}_{\ell},\mathcal{A}_{\ell}(\bm{m}_{\ell})\big),\\
 \mathcal{F}_{\ell,i}&=\big(\bm{m}_{\ell},\mathcal{A}_{\ell}(\bm{m}_{\ell}),\dots,\mathcal{B}_{\ell}^{i-1}(\mathcal{A}_{\ell}(\bm{m}_{\ell}))\big),\;i\geq 2,
 \end{split}
 \end{equation*}
where $\mathcal{A}_{\ell}$ and $\mathcal{B}_{\ell}$ are two blocks with the convolution $\bm{\kappa}$ and Rectified Linear Unit (ReLU) such that
\[\begin{split}
\mathcal{A}_{\ell}(\cdot):&=\text{ReLU}(\bm{\kappa}(\cdot)):\mathbb{R}^{n\times m}\rightarrow \mathbb{R}^{n\times m\times p},\\
\mathcal{B}_{\ell}(\cdot):&=\text{ReLU}(\bm{\kappa}(\cdot)):\mathbb{R}^{n\times m\times p}\rightarrow \mathbb{R}^{n\times m\times p},\\
\mathcal{B}_{\ell}^{i-1}&=\underbrace{\mathcal{B}_{\ell}\circ\dots\circ\mathcal{B}_{\ell}}_{i-1},\;\; \mathcal{G}_{\ell,i}=(\underbrace{\mathcal{A}_{\ell},\mathcal{B}_{\ell},\dots,\mathcal{B}_{\ell}}_{i}),\\
\mathcal{F}_{\ell,i}&=\text{Concat}(\mathcal{F}_{\ell,i-1},\mathcal{G}_{\ell,i}*\mathcal{F}_{\ell,i-1})\\
&=\text{Concat}\left(\mathcal{F}_{\ell,i-1},\sum\limits_{j=0}^{i-1}\mathcal{B}^j_{\ell}(\mathcal{A}_{\ell}(\bm{m}_{\ell}))\right),
\end{split}\]
where the convolution operation $\mathcal{G}_{\ell,i}$ of the layer $i$ is employed to recover the lost information of different sparse domains.
Finally, an $1\times1$ convolution $\mathcal{T}_{\ell}$ defined by
$\mathcal{T}_{\ell}=(\mathcal{T}^0_{\ell},\mathcal{T}^1_{\ell},\dots,\mathcal{T}^k_{\ell})$ is used to fuse features $\mathcal{F}_{\ell,k}$, i.e.
\[
\begin{split}
\bm{x}_{\ell}&=\mathcal{T}_{\ell}\ast\mathcal{F}_{\ell,k}=\mathcal{T}_{\ell}\ast\text{Concat}\left(\mathcal{F}_{\ell,k-1},\sum\limits_{j=0}^{k-1}\mathcal{B}^j_{\ell}(\mathcal{A}_{\ell}(\bm{m}_{\ell}))\right)\\
&=\mathcal{C}_{\ell}^0\ast\bm{m}_{\ell}+\sum_{j=1}^{k}\left(\mathcal{C}_{\ell}^j\ast\mathcal{B}^{j-1}_{\ell}(\mathcal{A}_{\ell}(\bm{m}_{\ell}))\right)=\sum_{j=0}^{k}\mathcal{C}^j_{\ell}\mathcal{G}_{\ell}^{j}(\bm{m}_{\ell}),
\end{split}\]
where each $1\times1$ convolution $\mathcal{C}^i_{\ell}$ in 
$\mathcal{C}_{\ell}=(\mathcal{C}^0_{\ell},\mathcal{C}^1_{\ell},\dots,\mathcal{C}^k_{\ell})$ is a linear combination of $\{\mathcal{T}^j_{\ell}\}_{j=0}^k$. The above operator $\mathcal{G}_{\ell}(\bm{m}_{\ell})$ embedded by many convolution blocks and ReLU layers can be learned as a more flexible representation of the non-linear operation $\mathcal{W}$ in (\ref{eq15}).

\textbf{Loss Function.}
The loss function is commonly employed to seek the real target image by minimizing the distance measure including pixel-based $\ell_1$-norm and $\ell_2$-norm between the final reconstructed output $\bm{x}_{\scriptscriptstyle\bm{f}}$ and the ground truth $\bm{x}$. However, the capability of pixel-based $\ell_2$-norm to capture perceptually relevant components, such as high-frequency geometric details, is insufficient because they are defined on basis of pixel-wise image differences \cite{KJiang2018}.

Here we adopt $\ell_1$-loss rather than $\ell_2$-loss to enlarge the original loss and add a constraint in the middle of iteration to make the optimization process more compliant. For the given training samples $\left\{\left(\bm{y}^{i}, \bm{x}^{i}\right)\right\}_{i=1}^{\mathcal{N}_s}$, the total loss is defined by
\begin{equation}
\mathcal{L}_{\mathrm{}}=\frac{1}{\mathcal{N}_s \mathcal{N}} \sum_{i=1}^{\mathcal{N}_s}\left(\left\|\bm{x}^{i}_{\scriptscriptstyle\overline{\mathcal{N}_{\ell}}}-\bm{x}^{i}\right\|_{1}+\left\|\bm{x}^i_{\scriptscriptstyle\bm{f}}-\bm{x}^{i}\right\|_{1}\right),
\label{eq35}
\end{equation}
where $\mathcal{N}_s$ is the sample number of training set, $\mathcal{N}$ is the size of $\bm{x}^{i}$, $\overline{\mathcal{N}_{\ell}}=\frac{\mathcal{N}_{\ell}+1}{2}$ and $\mathcal{N}_{\ell}$ is the stage number of our method.

\textbf{Parameters and Initialization.}
The learnable parameters of the proposed method are denoted as $\Theta=\left\{\eta_{\ell}, \mathcal{A}_{\ell},\mathcal{B}_{\ell},\mathcal{C}_{\ell}\right\}_{\ell=1}^{\mathcal{N}_{\ell}}$. The convolution network is initialized with Xavier algorithm. Moreover, $c_{1}$ and $c_{2}$ are initialized with $-0.2$ and $0.1$, respectively. $p$, $k$ and $\mathcal{N}_{\ell}$ are set to $32$, $8$ and $11$, respectively.

\begin{figure*}[htbp]
\centering
\includegraphics[width=0.95\textwidth]{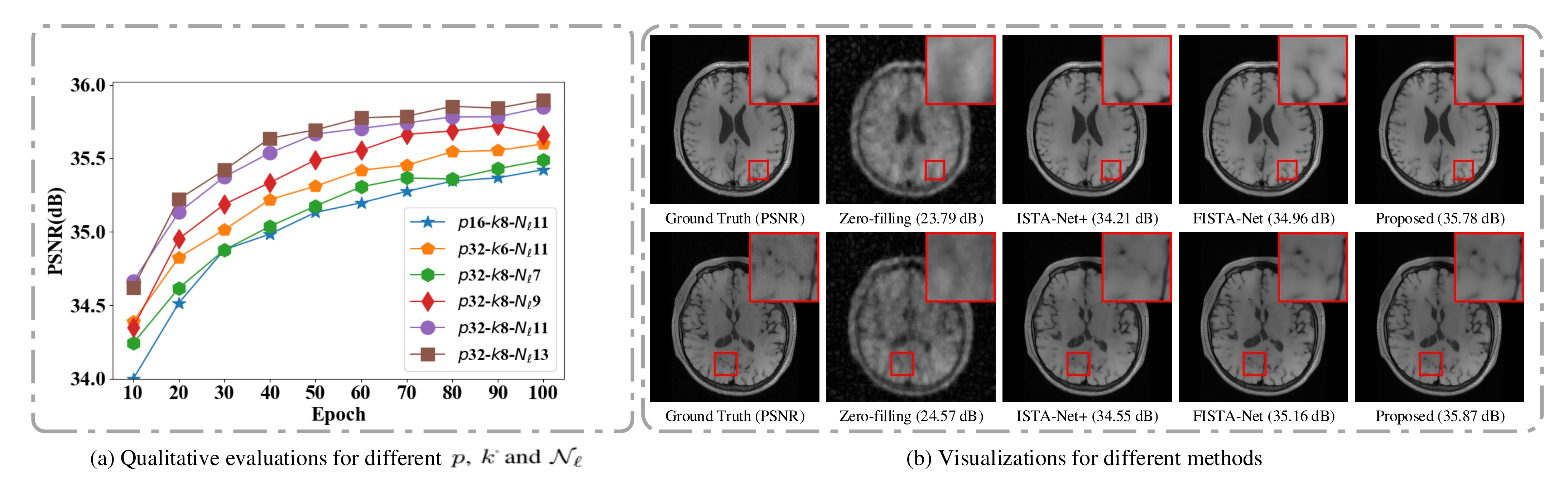}
\caption{Comparisons of different network parameters and different methods for brain MR images from CS Ratio $10\%$.}
\label{fig_results}
\end{figure*}

\section{Experiments and results}
We conducted a few experiments to compare the proposed methods with the state-of-the-art methods. Peak signal to noise ratio (PSNR) and Structural Similarity Index Measure (SSIM) are employed to evaluate their performances.

We evaluate the performances
of different methods on the widely used brain MR dataset \cite{Bermudez2020} 
for training within 800 MR brain medical images and testing within $50$ images as ISTA-Net+ \cite{Zhang2018}. They are T1-weighted 2D images from health and Alzheimer's disease individuals from different MRI devices. We denote the sampling matrix by $\mathcal{F}=PF$, where $P$ is an under-sampling matrix and $F$ is a discrete Fourier transform.

We use Pytorch to implement our method separately for different CS ratios with batch size $1$. We use Adam optimization \cite{Kingma2014} with a learning rate of $0.0001$ to train network for 100 epochs. All experiments are performed on a workstation with Intel Xeon CPU E5-2630 and Nvidia Tesla V100 GPU.

Firstly, we study the reconstruction performances of different basic network parameters ($p$, $k$ and $\mathcal{N}_{\ell}$) for brain MR images in CS Ratio $10\%$. As shown in Fig.\ref{fig_results}(a), a larger $p$, $k$ and $\mathcal{N}_{\ell}$ achieve quicker convergence and better reconstruction PSNR values. When $\mathcal{N}_{\ell}\geq11$, the improvement of reconstruction performance becomes slowly with increasing of $\mathcal{N}_{\ell}$. The proposed method allows deeper and wider networks without overfitting, and higher performance can be obtained by extracting more hierarchical features.

We compare the proposed method with Zero-filling \cite{Bernstein2001} and state-of-the-art unrolling methods (ISTA-Net+ \cite{Zhang2018}, FISTA-Net \cite{Xiang2020}). The CS reconstruction performances corresponding to five CS ratios are shown in Table \ref{table1}. FISTA-Net performs better than ISTA-Net+ in low CS ratios of 10\% and 20\%. Our method outperforms FISTA-Net and ISTA-Net+ in all CS ratios obviously.
\begin{table*}
\centering\newsavebox{\mybox}
\begin{lrbox}{\mybox}
\setlength{\tabcolsep}{2.4mm}{
\begin{tabular}{cccccccc}
\toprule[1.5pt]
\multirow{2}{*}{Index} & \multirow{2}{*}{Method} & \multicolumn{5}{c}{CS Ratio} \\ \cline{3-7}
& & 10\% & 20\% & 30\% & 40\% & 50\% \\
\midrule[0.95pt]
\multirow{4}{*}{PSNR} & Zero-filling & 26.64$\pm$3.79&	30.28$\pm$3.79&	32.89$\pm$3.84&	35.01$\pm$3.89&	36.92$\pm$3.93 \\
& ISTA-Net+ & 34.83$\pm$4.01	&38.75$\pm$3.86&	40.99$\pm$3.71&	42.64$\pm$3.64&	44.22$\pm$3.60 \\
& FISTA-Net & 35.15$\pm$3.88 &	38.84$\pm$3.78 &	40.99$\pm$3.75 &	42.61$\pm$3.65& 	44.03$\pm$3.47 \\
        & Proposed & \textbf{35.85$\pm$4.12}&	\textbf{39.26$\pm$3.87}&	\textbf{41.30$\pm$3.72}& 	\textbf{42.81$\pm$3.64}&	\textbf{44.32$\pm$3.57} \\
        \midrule[0.95pt]
        \multirow{4}{*}{SSIM} & Zero-filling& 0.5733$\pm$0.1335 &	0.6948$\pm$0.1198& 0.7736$\pm$0.1068 & 0.8268$\pm$0.0933 &0.8651$\pm$0.0802 \\
        & ISTA-Net+ & 0.9047$\pm$0.0489 &	0.9492$\pm$0.0240 &0.9639$\pm0.0167$	&0.9726$\pm$0.0127 &	0.9795$\pm$0.0097\\
        & FISTA-Net & 0.9115$\pm$0.0459& 	0.9502$\pm$0.0233 &0.9642$\pm$0.0169& 	0.9726$\pm$0.0128 &0.9792$\pm$0.0096 \\
        & Proposed & \textbf{0.9194$\pm$0.0452}& 	\textbf{0.9529$\pm$0.0226}& 	\textbf{0.9656$\pm$0.0160}& 	\textbf{0.9737$\pm$0.0124}& 	\textbf{0.9802$\pm$0.0095} \\
         \bottomrule[1.5pt]
    \end{tabular}}
\end{lrbox}
\caption{Average performance comparisons with different CS ratios for CS-MRI.}
\scalebox{0.95}{\usebox{\mybox}}\label{table1}
\end{table*}

In Fig.\ref{fig_results}(b), we show the brain MRI reconstruction results of these methods in CS ratio of $10\%$, which are consist of total image with a zoom-in details. Zero-filling can remove some noise, but it does not recover the rich texture information very well. Although FISTA-Net and ISTA-Net+ are good at reconstructing MRI from CS measurement, the details are still not good enough. The proposed method can achieve more better reconstruction performance and recover more geometric characteristics.

\section{Conclusions}
In this study, we focus on the classical model-based CS-MRI optimization problem and propose a novel deep geometric distillation CS-MRI reconstruction network. The proposed method combines the merits of model-based CS-MRI method with sampling constraints and deep learning-based CS-MRI method with powerful learning ability. We unfold CS-MRI optimization problem and design deep compensation module to extract lost texture details in the linear reconstruction stage from different geometric characteristic domains. This combination can give us a new perspective to design the explainable network. Experimental validations on the public brain
datasets demonstrate the potential of the proposed network both in terms
of qualitative and quantitative results.

\section*{Acknowledgment}
This work was supported by the National Natural Science Foundation of
China (NSFC Project no.11771369, no.12071402 and no.11931003), also
partly by grants from the outstanding young scholars of Education Bureau
of Hunan Province, P. R. China (number 17B257) and Natural Science
Foundation of Hunan Province, P. R. China (2018JJ2375, 2018XK2304,
2020JJ2027 and 2018WK4006) and Hunan Provincial Innovation Foundation
For Postgraduate (number XDCX2021B097).

\bibliographystyle{hsiam} 
\bibliography{ref}






\end{document}